\documentclass[12pt]{iopart}
\usepackage{graphicx, subfigure}
\begin{document}

\title[Charged Particle Production at Large Transverse Momentum]{Charged Particle Production at Large Transverse Momentum in Pb$-$Pb Collisions at $\sqrt{s_{NN}}=2.76$ TeV Measured with ALICE at the LHC.}
\author{Jacek Otwinowski for the ALICE Collaboration}

\address{GSI Helmholtz Centre for Heavy Ion Research GmbH, Planckstrasse 1, 64291~Darmstadt, Germany}
%\ead{j.otwinowski@gsi.de}

\begin{abstract}
Transverse momentum  ($p_{T}$) spectra of charged particles are measured as a function of event centrality in Pb$-$Pb collisions at  $\sqrt{s_{NN}}=2.76$~TeV with ALICE at the LHC. The spectra are compared to those measured in pp collisions at the same collision energy in terms of the nuclear modification factor $R_{AA}$. The high-$p_{T}$ charge particle production in central Pb$-$Pb collisions ($0-5\%$) is strongly suppressed by a factor $\approx6$ at transverse momenta $p_{T}=6-7$~GeV/c as compared to expectation from independent superposition of nucleon-nucleon collisions. Above $p_{T}=7$~GeV/c there is a significant rise in the nuclear modification factor, which reaches $R_{AA} \approx 0.4$ at $p_{T}=50$~GeV/c. The measured suppression of high-$p_{T}$ particles is stronger than that measured at RHIC.

\end{abstract}

%\maketitle
\section{Introduction}

This paper reports on measurements of charged particle yields as a function of transverse momentum and event centrality in Pb$-$Pb collisions at $\sqrt{s_{NN}}=2.76$~TeV recorded  by ALICE  {\cite{ALICE_DET}} in November 2010. 

The measurement is motivated by results \cite{BRAHMS_RAA,PHOBOS_RAA,STAR_RAA,PHENIX_RAA} from the Relativistic Heavy Ion Collider (RHIC), which showed that hadron production at large transverse momentum in central Au$-$Au collisions at $\sqrt{s_{NN}}=200$~GeV is suppressed by a factor $4-5$ compared to expectations from an independent superposition of nucleon-nucleon collisions. This observation is typically expressed in terms of the nuclear modification factor which is defined as the ratio of the charged particle yield in Pb$-$Pb to that in pp (pp reference), scaled by the number of binary nucleon-nucleon collisions $\langle N_{coll} \rangle$

\begin{equation}\label{EQUATION_RAA}
R_{AA}(p_{T}) = \frac{(1/N^{AA}_{evt})d^{2}N^{AA}_{ch}/d\eta dp_{T}}{\langle N_{coll} \rangle (1/N^{pp}_{evt})d^{2}N^{pp}_{ch}/d\eta dp_{T}}
\end{equation}
In absence of nuclear modifications $R_{AA}$ is unity at high $p_{T}$. 
 
 The following analysis is based on $21 \times 10^{6}$ minimum-bias Pb$-$Pb events. The details of the analysis can be found in {\cite{ALICE_MULT, ALICE_RAA}}. For selected tracks the transverse momentum resolution is $\sigma(p_{T})/p_{T}\approx10$~$\%$ at $p_{T}=50$~GeV/c.
 
\begin{figure}[t]
\begin{center}$
\begin{array}{cc}
\includegraphics[width=7.5cm,height=8.5cm]{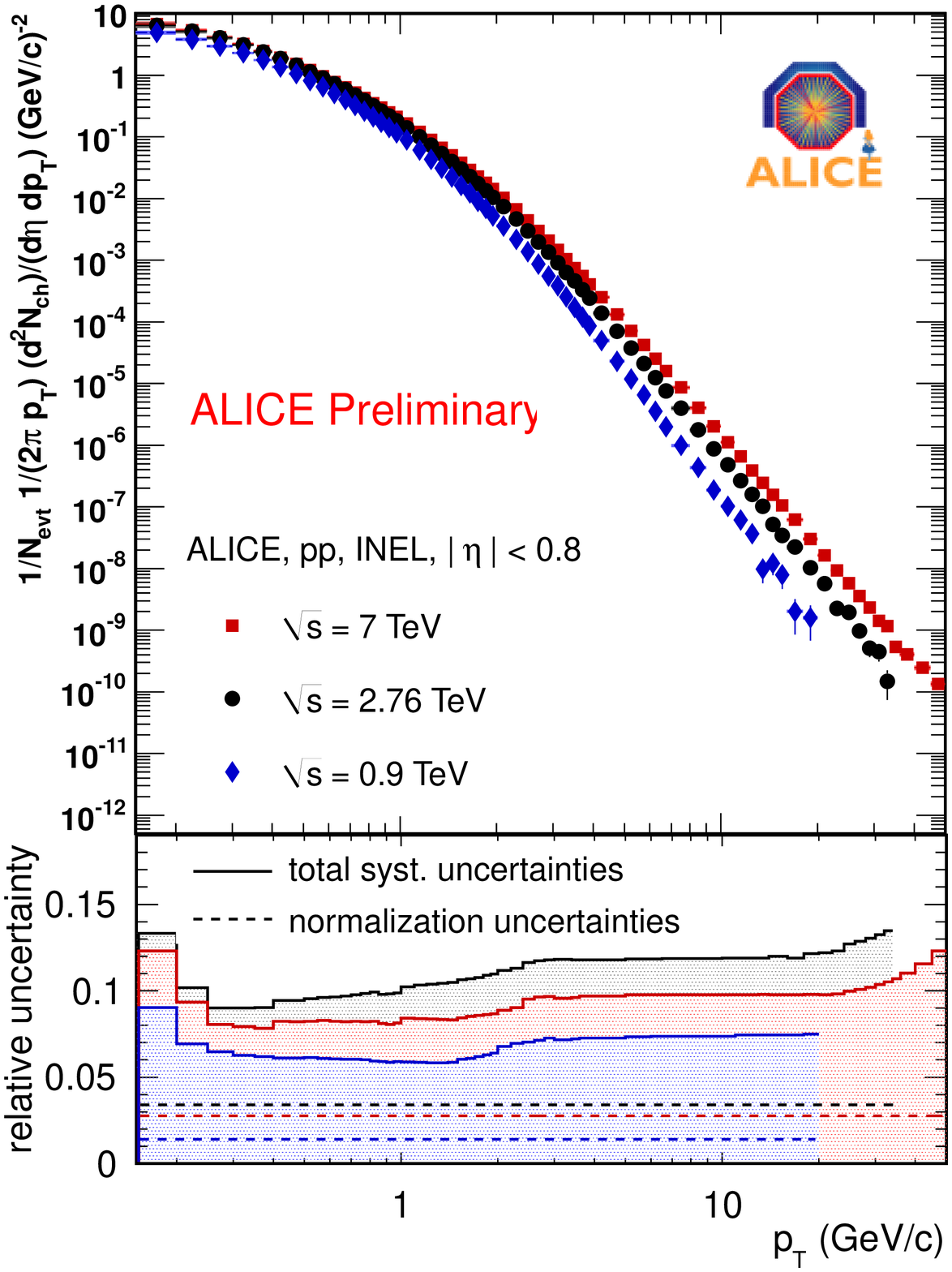} &
\includegraphics[width=7.5cm,height=8.5cm]{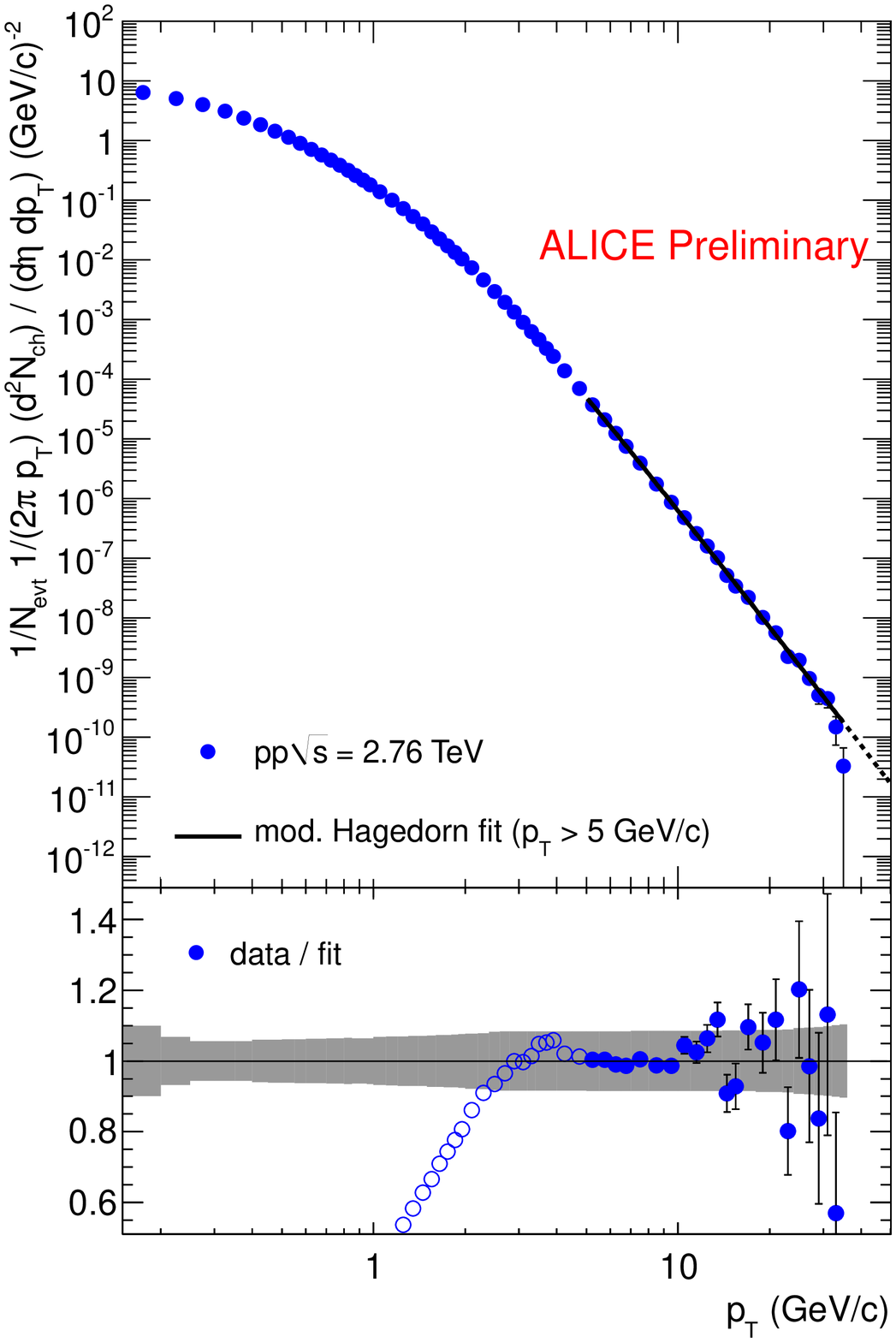}
\end{array}$
\end{center}
\caption{\label{PPPTSPECRA} {\bf Left:} $p_{T}$ spectra of charged particles measured in pp collisions at  $\sqrt{s}=0.9$, 2.76 and 7 TeV. The systematic uncertainties are shown in the lower panel. {\bf Right:} $p_{T}$ spectrum of pp collisions at  $\sqrt{s}=2.76$~TeV together with the parametrization. For $p_{T}>30$~GeV/c the extrapolation is shown. The lower panel shows the data--to--fit ratio where the grey area denotes the $p_{T}-$dependent systematic uncertainties.}
\end{figure}

\section{Results}

\begin{figure}[t]
\begin{center}$
\begin{array}{cc}
\includegraphics[width=7.5cm,height=6.5cm]{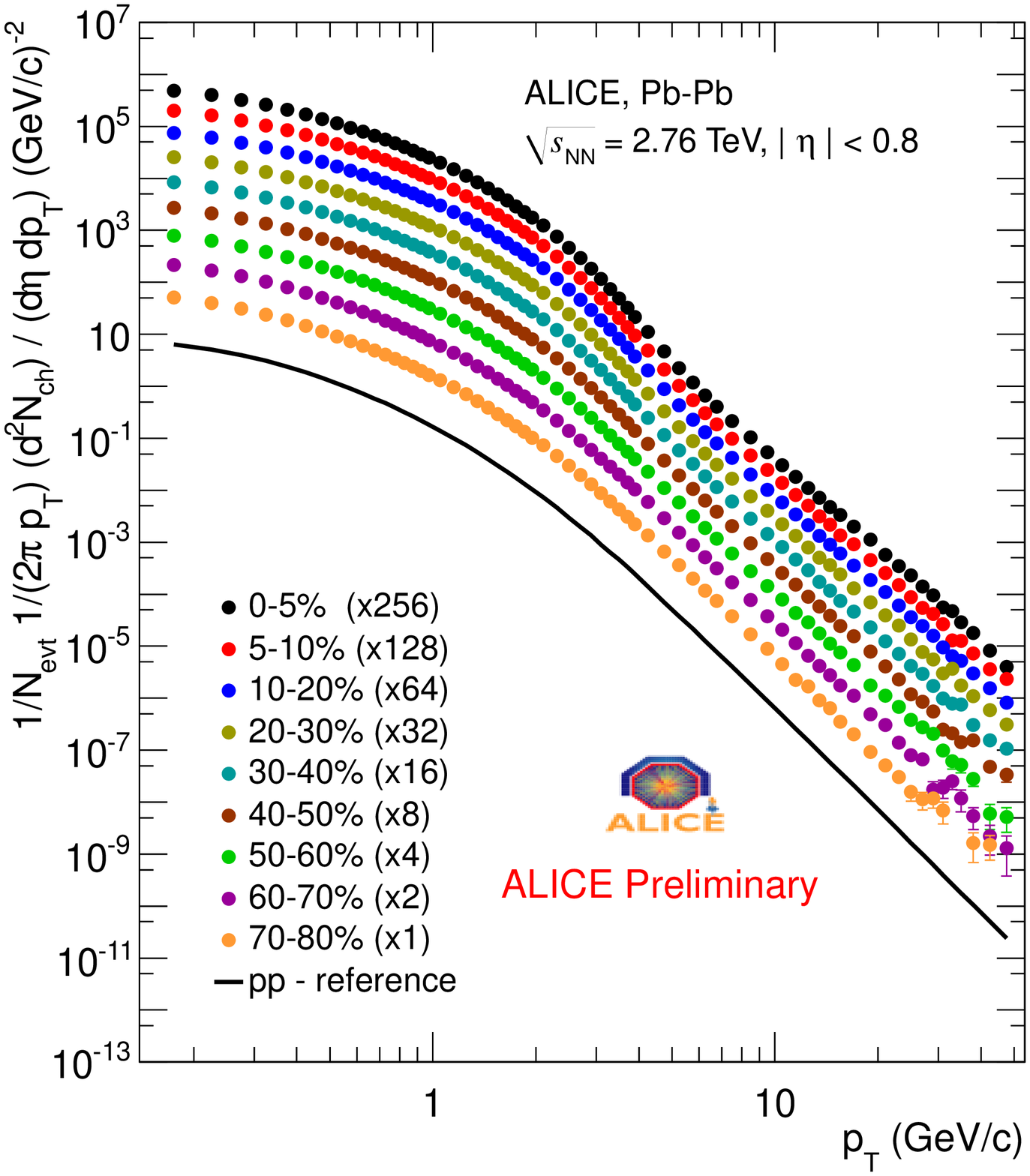} &
\includegraphics[width=7.8cm,height=6.6cm]{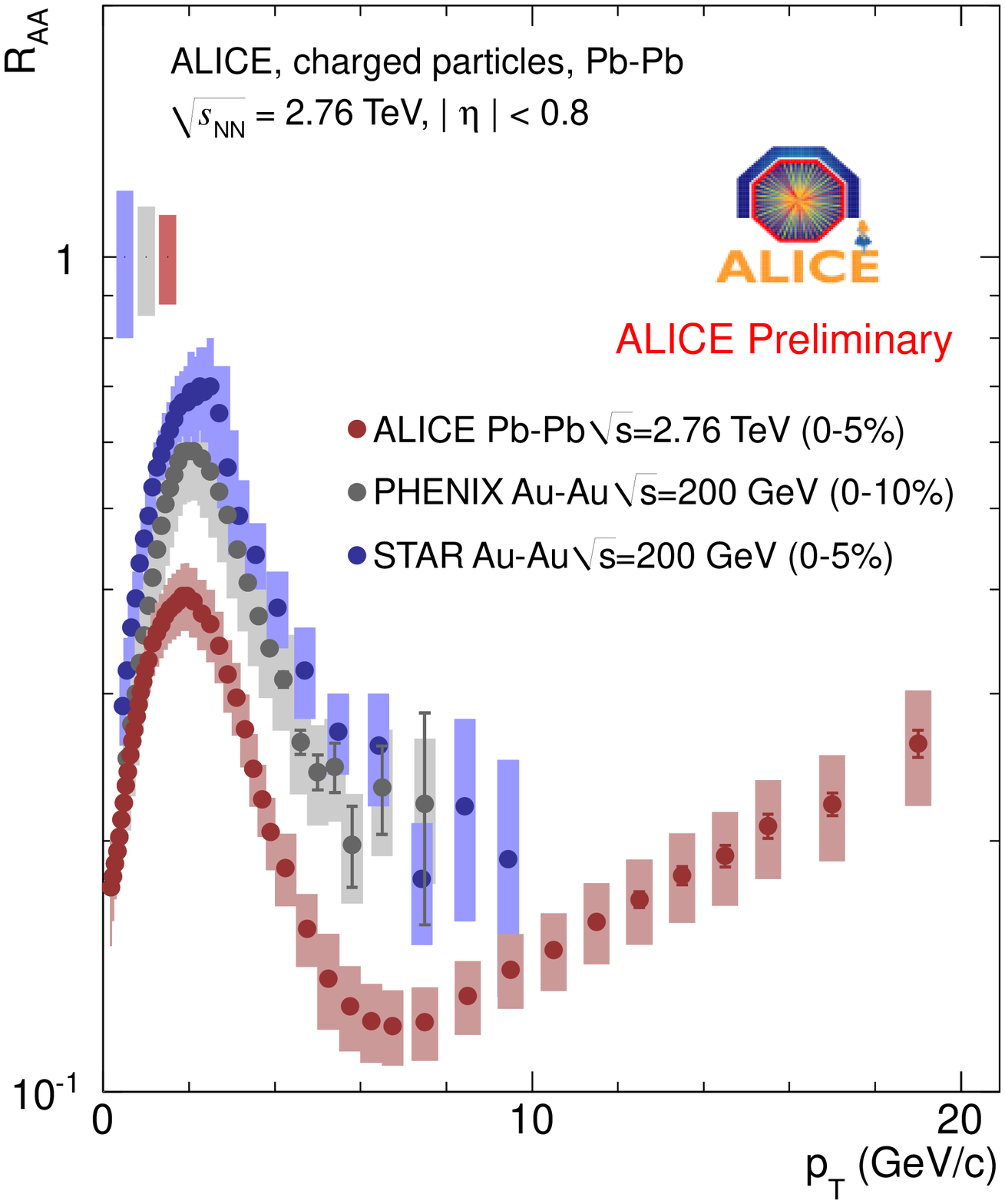}
\end{array}$
\end{center}
\caption{\label{PbPbPTSPECRA} {\bf Left:}  $p_{T}$  of charged particles measured in Pb$-$Pb collisions at $\sqrt{s_{NN}}=2.76$~TeV in different centrality intervals. For better visibility the spectra are separated with scaling factors indicated in the plot.The systematic and statistical uncertainties are added in quadrature. The solid line denotes the pp reference spectrum (details in the text). {\bf Right:} $R_{AA}$ of primary charged particles measured with ALICE in central Pb$-$Pb collisions ($0-5\%$) in comparison to RHIC measurements. The error bars at $R_{AA}$=1 denote the contributions from normalization uncertainties.}        
\end{figure}

Fig. \ref{PPPTSPECRA} (left panel) shows fully corrected $p_{T}$ spectra of unidentified charged particles measured in pp collisions at $\sqrt{s}= 0.9$, 2.76 and 7 TeV. The distributions reveal the typical power law shape at high $p_{T}$, characteristic for hard scattering. The $p_{T}$ spectrum flattens with higher pp collision energy. Due to the limited statistics of the $\sqrt{s}=2.76$~TeV data set the pp reference for $R_{AA}$ was constructed using the measured yield only below $p_{T}=5$~GeV/c, as indicated in Fig. \ref{PPPTSPECRA} (right panel). At larger $p_{T}$ the pp reference was approximated by a modified Hagedorn function. This functional form provides the best fit to the measured pp spectrum between $p_{T}=5$ and 30 GeV/c and was extrapolated to $p_{T}=50$ GeV/c. The estimated uncertainty on the pp reference due to the parametrization and extrapolation procedure increases with $p_{T}$ and reaches $25\%$ at $p_{T}=50$ GeV/c.

Fig.  \ref{PbPbPTSPECRA} (left panel) shows the fully corrected $p_{T}$ spectra of unidentified charged particles measured in Pb$-$Pb collisions at $\sqrt{s_{NN}}=2.76$~TeV in different centrality intervals. 
In the most peripheral collisions the shape of the $p_{T}$ distribution is similar to pp. In contrast, a marked depletion of the spectra is developing gradually as centrality is increasing, indicating a significant suppression of particle production in central collisions. 

In Fig. \ref{PbPbPTSPECRA} (right panel) $R_{AA}$ measured with ALICE is compared to measurements at lower collision energy ($\sqrt{s_{NN}} = 200$~GeV) by the PHENIX and STAR experiments {\cite{PHENIX_RAA_2, STAR_RAA_2}} at RHIC. At $p_{T}=1$~GeV/c the measured value of $R_{AA}$ is similar to those of RHIC. The position and shape of the maximum at $p_{T}\approx2$ GeV/c and the subsequent decrease are similar at RHIC and LHC. At $p_{T} $= 6--7~GeV/c  $R_{AA}$ is smaller than at RHIC indicating that a very dense medium is formed in Pb$-$Pb collisions at the LHC.

\begin{figure}[t]
\begin{center}$
\begin{array}{cc}
\includegraphics[width=3.0in]{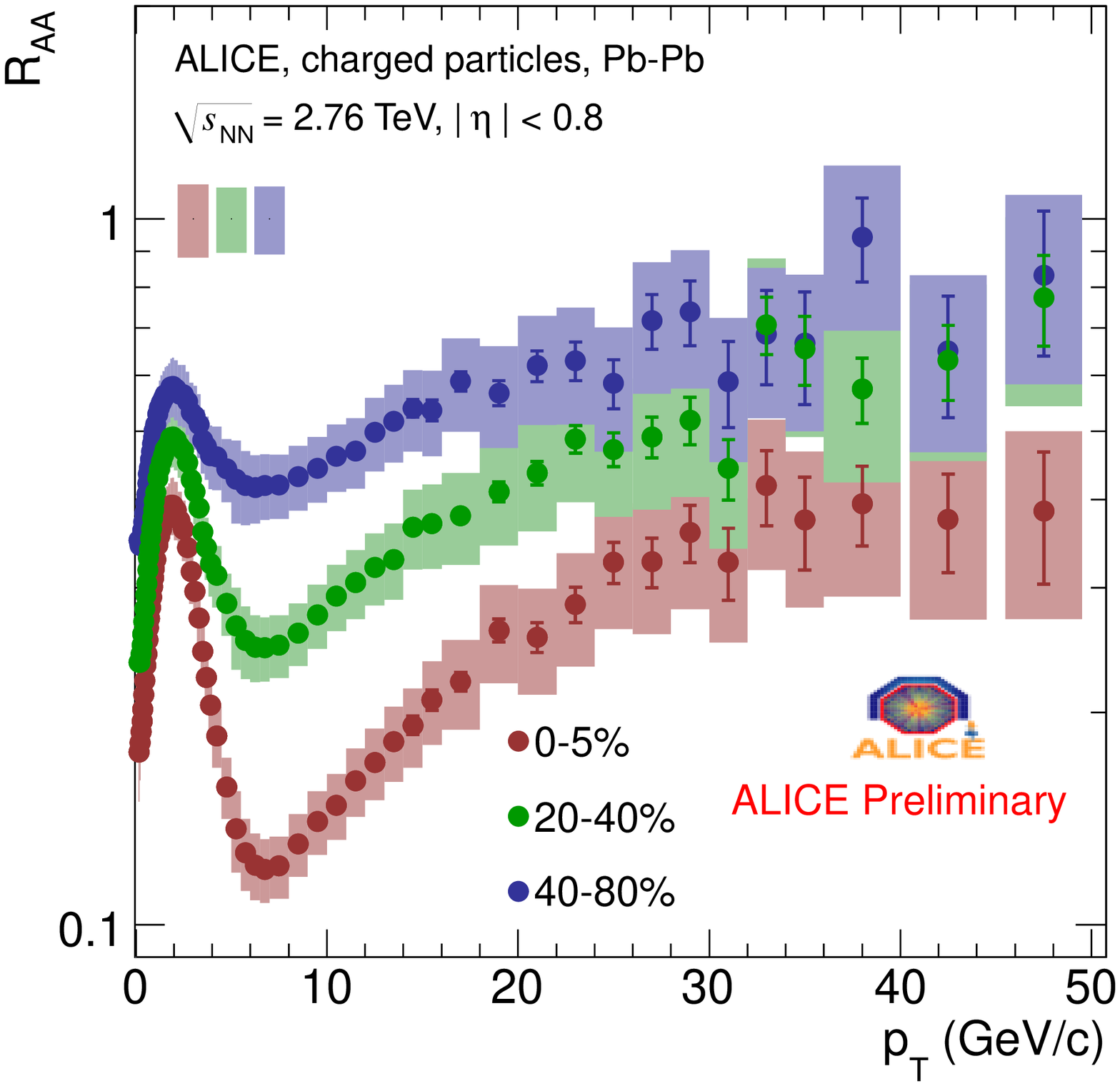} &
\includegraphics[width=3.0in]{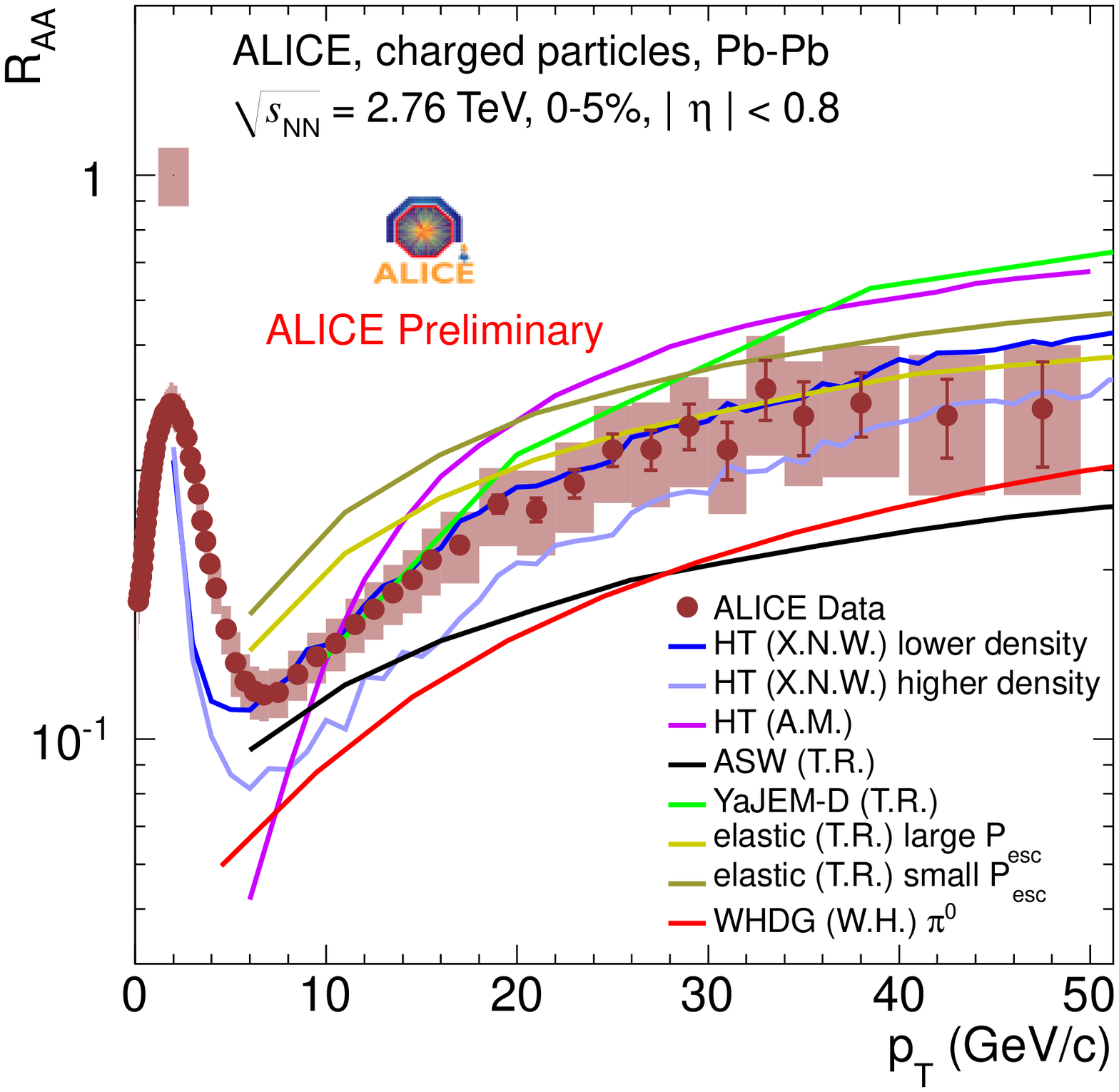}
\end{array}$
\end{center}
\caption{\label{RAASPECTRA} {\bf Left:}   $R_{AA}$ of charged particles measured with ALICE in central Pb$-$Pb collisions in three centrality intervals. {\bf Right:} $R_{AA}$ of charged particles measured with ALICE in central Pb$-$Pb collisions ($0-5\%$) in comparison to model calculations. The error bars at $R_{AA}$=1 denote contributions from normalization uncertainties.}
\end{figure}

The nuclear modification factors out to  $p_{T}=50$~GeV/c are shown in Fig. {\ref{RAASPECTRA}} (left panel, \footnote{{\bf After the conference presentation, we found small inconsistencies in the tracking, leading to an overestimation of the $R_{AA}$ at high transverse momenta. The analysis was reviewed, and here we present the updated result in the transverse momentum range $p_{T}< 50$~ GeV/c, where the updated values remain within the systematic uncertainties of the ones shown at the conference.}}) for different centrality intervals. At all centralities, a pronounced minimum at
about $p_{T}=6-7$~GeV/c is observed. For $p_{T}>7$~GeV/c there is a significant rise in the nuclear modification factor until $p_{T}=30$~GeV/c.  This emphasizes the strong relation between the medium density and partonic energy loss. 

Fig. {\ref{RAASPECTRA}} (right panel) shows a comparison of $R_{AA}$ in central Pb$-$Pb collisions  ($0-5\%$) at  $\sqrt{s_{NN}}$=2.76~TeV to calculations from energy loss models {\cite{TH_1,TH_2, TH_3,TH_4}}. All model
calculations have been constrained to match $R_{AA}$ results from RHIC. The qualitative features of our data are described by all models, including the strong
rise of $R_{AA}$ below $p_{T}=30$~GeV/c and the flattening off at higher $p_{T}$. A more quantitative comparison of model calculations to the present data will help to put tighter constraints on the underlying energy loss mechanisms. 

\section{Summary}

The results indicate a strong suppression of charged particle production in Pb$-$Pb
collisions at $\sqrt{s}=2.76$~TeV and a characteristic centrality and $p_{T}$ dependence of the nuclear modification factors. The suppression observed in central Pb$-$Pb collisions at the LHC is
stronger than in central Au$-$Au collisions at RHIC. The comparison of ALICE data to model calculations indicates a large sensitivity of high-$p_{T}$ particle production to details of energy loss mechanisms.

\end{document}